\documentstyle[12pt,a41,epsfig,cite]{article}
\newcommand{\ve}{\varepsilon}

\begin{document}
\sloppy
\thispagestyle{empty}
\begin{flushleft}
DESY 04--120    \hfill
{\tt hep-ph/0411111}\\
NIKHEF--2004--014\\
SFB/CPP--04--28\\
June 2004  \\
\end{flushleft}

\mbox{}
\vspace*{\fill}
\begin{center}
 
{\Large\bf The 16th Moment of the Non--Singlet Structure}

\vspace{2mm}
{\Large \bf
\boldmath Functions $F_2(x,Q^2)$ and $F_L(x,Q^2)$ to $O(\alpha_s^3)$}

\vspace{4em}
\large
J. Bl\"umlein$^a$ and J.A.M. Vermaseren$^b$
\normalsize
\\
\vspace{4em}
{\it$^a$DESY Zeuthen}\\
{\it    Platanenallee 6, D--15738 Zeuthen, Germany}\\

\vspace{4mm}
{\it $^b$NIKHEF Theory Group}\\
{\it Kruislaan 409, 1098 SJ Amsterdam, The Netherlands}\\
\end{center}
\vspace*{\fill}
\begin{abstract}
\noindent
We present the results of an analytic next--to--next--to leading order QCD
calculation of the non--singlet anomalous dimension $\gamma_{\rm NS}^+(N)$
and the coefficient functions $C_{2,L}(N)$ associated to the deeply inelastic
structure functions $F_2(x,Q^2)$ and $F_L(x,Q^2)$ for the Mellin moment $N=16$.
Comparisons are made with results in the literature.
\end{abstract}
\vspace*{\fill}
\newpage
\section{Introduction}

\vspace{1mm}
\noindent
One of the most precise methods to determine the strong coupling 
constant $\alpha_s(Q^2)$ is to measure it from the scaling violations 
of the structure functions in deeply inelastic scattering. The present 
level of experimental accuracy is of  order  $\delta 
\alpha_s(M_Z^2) \simeq \pm 0.001$ \cite{H1Z}. Since the theory errors due to 
factorization and renormalization scale uncertainties are of $\delta 
\alpha_s(M_Z^2)_{\rm sc} \simeq \pm 0.005$ \cite{NLOPR} in next--to--leading order
the knowledge of the 3--loop anomalous dimensions is required to reduce 
this error to the level reached by experiment.\footnote{Cf.~Ref.~\cite{BBG,BETH} for recent 
comparisons of the measured value of $\alpha_s(M_Z^2)$.}  

The one--loop anomalous dimensions were calculated in \cite{gLO} and the one--loop 
coefficient functions were obtained in a series of subsequent calculations, see 
Ref.~\cite{FP} for a 
summary of the results. Later the two--loop anomalous dimensions 
\cite{gNLO} and the two--loop coefficient functions 
\cite{cNLO} were evaluated applying different methods. The calculation of the 
anomalous dimensions and Wilson coefficients in three loop order is a very difficult task.
The first calculations were performed for a series of sum rules and
the individual moments for $N = 2, 4, 6, 8$~\cite{NNLOmom} in the non--singlet and singlet 
case as well as for $N = 10$ in the non--singlet case 
using the symbolic 
manipulation program {\tt FORM}~\cite{FORM}. With growing 
computational power the moments $N = 10, 12$ in the singlet case and $N = 12, 14$ in the 
non--singlet case as well as the moments $N = 3, 5, 7, 9, 11, 13$  associated to the 
structure function $xF_3(x,Q^2)$ could be calculated \cite{MOM14}. Very 
recently, the
3--loop non--singlet \cite{3lns} and singlet \cite{3lsin} anomalous dimensions were 
calculated for deep--inelastic scattering off unpolarized nucleons.

In this paper we calculate the 16th moment of the non--singlet splitting function,
$\gamma_{\rm NS}^{(16),+}(N, Q^2)$, and the Wilson coefficients of the structure functions
$F_2(x,Q^2)$ and $F_L(x,Q^2)$, $C^{2,L}_{\rm NS}(N, Q^2)$, for the case of pure photon 
exchange to 3--loop order using the {\tt MINCER} formalism \cite{MINCER}.
This calculation was started before complete 3--loop results became available and serves as an 
independent check of the result for the non--singlet 
anomalous dimension given in Ref.~\cite{3lns} and
predicts the 16th moment  for the non--singlet coefficient function,   
\cite{3lFL,3lcoef}.
The paper is organized as follows. In section~2 we give an outline of the basic 
formalism. The renormalization of the hadronic matrix element is described in section~3.
The results of the calculation are given in section~4 and section~5 contains the 
conclusions.   
\section{Basic formalism}
The hadronic tensor for  deeply inelastic scattering in the case of pure photon exchange is 
given by ~:
\begin{eqnarray}
W_{\mu \nu}(x,Q^2) &=& \frac{1}{2\pi} {\sf Im}~~T_{\mu 
\nu}(p,q)\nonumber\\
&=& \frac{1}{4\pi} \int d^4 x {\rm e}^{{\rm i}qx} \langle P|J_{\mu}(x) 
J_{\nu}(0)|
P\rangle \nonumber\\
&=& - \left[g_{\mu \nu} + \frac{2x}{q^2} \left(p_\mu q_\nu + p_\nu q_\mu\right) + 
\frac{4x^2}{q^2} p_\mu p_\nu \right] \frac{1}{2x}~F_2(x,Q^2) 
\nonumber\\ & & + \left(g_{\mu \nu} 
-
\frac{q_\mu q_\nu}{q^2}\right) \frac{1}{2x}~F_L(x,Q^2)~.
\end{eqnarray}
Here $x = Q^2/(2 p.q)$ denotes the Bjorken variable, $P$ is the nucleon momentum, $q$ denotes 
the 
space--like momentum transfer from the leptons to 
the nucleon with $q^2 =- Q^2$ and the 
Compton amplitude reads
\begin{eqnarray}
\label{eqTT0}
T_{\mu \nu}(p,q) = {\rm i}~\int~d^4 z {\rm e}^{{\rm i}qz} \langle 
P|T\left[J_\mu(z) J_\nu(0)\right]|P\rangle~.
\end{eqnarray}
We now consider the light--cone expansion of the (forward) Compton 
amplitude~\cite{LCE}
\begin{eqnarray}
\label{eqTT}
{\rm i}~\int~d^4 z {\rm e}^{{\rm i}qz} \left.
T\left[J_{\nu_1}(z) J_{\nu_2}(0)\right]\right|_{\rm NS}
&=& 
\sum_{N} \left(\frac{1}{Q^2}\right)^N 
\Biggl[
\Biggl( g_{\nu_1 \nu_2} - \frac{q_{\nu_1} q_{\nu_2}}{q^2}\Biggr) 
q_{\mu_1} q_{\mu_2} C_{L,N}^{\rm NS} \left(\frac{Q^2}{\mu^2},a_s \right)
\nonumber\\
& & - \Biggl(g_{\nu_1 \mu_1} g_{\mu_2 \nu_2} q^2 - g_{\nu_1 \mu_1} 
q_{\nu_2} q_{\mu_2} - g_{\nu_2 \mu_2} q_{\nu_1} q_{\mu_1} + g_{\nu_1 
\nu_2}
q_{\mu_1} q_{\mu_2} \Biggr) \nonumber \\ 
& & \times C_{2,N}^{\rm NS} \left(\frac{Q^2}{\mu^2}, a_s \right) 
\Biggr] 
q_{\mu_3} \ldots q_{\mu_N} O^{{\rm NS},~\{\mu_1, \ldots, \mu_N\}}(0)
\nonumber\\ & &
+~~{\rm higher~~twists}~.
\end{eqnarray}
The twist--2 contributions to the local non--singlet operators $O^{\rm NS}(0)$ are
linear combinations out of the operators 
\begin{eqnarray}
O^{\alpha,~{\mu_1, \ldots, \mu_N}} &=& 
\overline{\psi} \lambda^\alpha
\gamma^{\{\mu_1} D^{\mu_2} \cdots D^{\mu_N\}} \psi,~~~\alpha = 1,2, \ldots, 
(N_F^2 -1)~, 
\end{eqnarray}
$\mu^2$ denotes the factorization 
scale and $a_s = \alpha_s/(4\pi) = g^2/(4 \pi)^2$, with $g$  the strong coupling 
constant. The functions $C_{(L,2),N}^{\alpha}$ are the Wilson coefficients associated to the
moment index $N$. 
The Mellin moments of the structure functions $F^{\rm NS}_{2,L}(x,Q^2)$ 
are given by
\begin{eqnarray}
M_{k,N-2} = \int_0^1 dx~x^{N-2}~F_k^{\rm NS}(x,Q^2)
= C_{k,N}^{\rm NS} \left(\frac{Q^2}{\mu^2}, a_s 
\right) A_{{\rm nucl},N}^{\rm NS},~~~~k = 2, L~.
\end{eqnarray}
For pure electromagnetic interactions $T_{\mu \nu}$ is even in $x$. The 
corresponding crossing relation has the consequence that only 
the even moments contribute to (\ref{eqTT}). The nucleon matrix elements 
are
\begin{eqnarray}
\langle P|O^{{\rm NS},~\{\mu_1, \ldots, \mu_N\}}|P\rangle = P^{\{\mu_1 \ldots}
P^{\mu_N\}} A^{\rm NS}_{{\rm nucl}, N}\left(\frac{P^2}{\mu^2}\right)~.
\end{eqnarray}
The scale--dependence of the non--singlet coefficient function is governed
by the renormalization group~:
\begin{eqnarray}
\left[
\mu^2 \frac{\partial}{\partial \mu^2} + \beta(a_s) 
\frac{\partial}{\partial 
a_s} - \gamma^{\rm NS}_N \right] C_{i,N}^{\rm NS}\left(\frac{Q^2}{\mu^2}, 
a_s\right) = 0,~~~~~~i=2, L~.
\end{eqnarray}  
The $\beta$--function rules the scale dependence of the coupling constant
$a_s(\mu^2)$
\begin{eqnarray}
\label{asrun}
\mu^2 \frac{\partial a_s(\mu^2)}{\partial \mu^2} = \beta(a_s) = 
- \sum_{l=0}^{\infty} \beta_l~a_s^{l+2}(\mu^2) 
\end{eqnarray}
on the renormalization scale $\mu = \mu_R$, where $\beta_l$ are the expansion 
coefficients. 
To 3--loop order the
constants $\beta_l$ for $SU(N)$~\cite{BET} 
\begin{eqnarray}
\beta_0 &=& \frac{11}{3} C_A - \frac{4}{3} T_F N_F \\
\beta_1 &=& \frac{34}{3} C_A^2 - 4 C_F T_F N_F - \frac{20}{3} C_A T_F 
N_F \\
\beta_2 &=& \frac{2857}{54} C_A^3 + 2 C_F^2 T_F N_F - \frac{205}{9} C_F 
C_A T_F N_F \nonumber\\
& & - \frac{1415}{27} C_A^2 T_F N_F + \frac{44}{9} C_F T_F^2 N_F^2 +
\frac{158}{27} C_A T_F^2 N_F^2  
\end{eqnarray}
contribute, with $C_A = N_c, C_F = (N_c^2 - 1)/(2 N_c)$,  and
$T_F = 1/2$ and $N_c = 3$. $N_F$ denotes the number of flavors.
In the QCD--improved parton model the forward Compton amplitude (\ref{eqTT0}), valid for 
nucleon
states $|P\rangle$, reduces to that for photon--quark scattering, since only one initial state 
quark participates in the scattering process, with $p$ the quark 4--momentum,
\begin{eqnarray}
\label{eqTT1}
T_{\mu \nu}^{q \gamma q \gamma} = {\rm i}~\int~d^4 z~{\rm e}^{iqz}~\langle 
p|T\left[J_{\mu}(z) J_{\nu}(0)\right]|p \rangle~.
\end{eqnarray}
One applies the projector
\begin{eqnarray}
P_N \equiv  
\left[
\frac{q^{\{\mu_1 \ldots \mu_N \}} }{N!} 
\frac{\partial^N}{\partial p^{\mu_1} \cdots \partial p^{\mu_N}} 
\right]_{p^2=0} 
\end{eqnarray}
onto the Mellin moments $N$ and the Lorentz projectors 
\begin{eqnarray}
P_L^{\mu \nu} &=& - \frac{q^2}{(p.q)^2} p^{\mu} p^\nu \\  
P_2^{\mu \nu} &=& - \left(\frac{3 - 2 \varepsilon}{2(1 - \varepsilon)} 
\frac{q^2}{(p.q)^2} p^\mu p^\nu + \frac{1}{2(1 - \varepsilon)} g^{\mu 
\nu}\right)~,
\end{eqnarray}
valid in $D = 4 - 2 \varepsilon$ dimensions. These projections lead to the
following moments of the Compton amplitude
\begin{eqnarray}
\label{eq19}
T_{k,N}^{q \gamma q \gamma}\left(\frac{Q^2}{\mu^2}, a_s, 
\varepsilon \right) = C_{k,N}^{\rm NS}\left(\frac{Q^2}{\mu^2}, a_s, 
\varepsilon \right) Z^{\rm NS}_N \left(a_s, \frac{1}{\varepsilon} \right) 
A_{q,N}^{\rm NS, tree}(\varepsilon),~~~~~k = 2, L~.
\end{eqnarray}

\section{Renormalization}
\label{secren}

\vspace{1mm}
\noindent
The coefficient functions $C$ and $Z$--factors in (\ref{eq19})
obey the following representations~:
\begin{eqnarray}
\label{eqC}
C(a_0,\ve) &=& \delta + a_0 \left(C_{10} + \ve C_{11} + \ve^2 C_{12}\right)
             + a^2_0 \left (C_{20} + \ve C_{21}\right) + a^3_0 C_{30} + O(a_0^4)~, 
\\
\label{eqZ}
Z(a,\ve) &=& 1 + a_0~\frac{Z_{11}}{\ve}+ a_0^2~\left(\frac{Z_{22}}{\ve^2} + 
\frac{Z_{21}}{\ve}\right) + a^3_0~\left(\frac{Z_{33}}{\ve^3} +
\frac{Z_{32}}{\ve^2} + \frac{Z_{31}}{\ve} \right) + O(a^4_0)~,
\end{eqnarray}
with $\delta = 1$ for $C_2$ and $\delta = 0$ for $C_L$.
Here, $a_0$ denotes the bare coupling constant, which is related to the running coupling
by
\begin{eqnarray}
\label{eqa0}
a_0 = a - \frac{\beta_0}{\ve} a^2 + \left(\frac{\beta_0^2}{\ve^2} -
\frac{\beta_1}{2\ve}\right) a^3 + O(a^4)~, 
\end{eqnarray}
cf.~(\ref{asrun}). We identified the scales $\mu^2 = Q^2$.
Yet a separation between the 
contributions to the anomalous dimension and the coefficient functions is 
possible as outlined in the following. 
The anomalous dimension and the Wilson coefficient  are 
\begin{eqnarray}
\gamma(a,N) &=& \sum_{k=0}^{\infty} a^{k+1} \gamma_k \\
c(a,N)      &=& \delta + \sum_{k=1}^{\infty} a^{k} c_{k0}~.
\end{eqnarray}
The respective coefficients $\gamma_k$ and $c_{k0}$ are determined as
follows. We denote by $C(\xi)$ the coefficient of type $\xi$ in $T$
(\ref{eq19}). Identifying the corresponding powers in $\ve$ one obtains to $O(a^3)$~:
\begin{eqnarray}
\gamma_0 &=& C\left(\frac{a}{\ve}\right) \\
\gamma_1 &=& 2\left[C\left(\frac{a^2}{\ve}\right) - \gamma_0
             c_{10}\right] \\
\gamma_2 &=& 3\left[C\left(\frac{a^3}{\ve}\right) - c_{10} Z_{21} 
             - C_{11} Z_{22} - \gamma_0 c_{20}\right]\\
c_{10}   &=& C\left(a \ve^0\right) \\
c_{20}   &=& C\left(a^2 \ve^0\right) - \gamma_0 C_{11} \\
c_{30}   &=& C\left(a^3 \ve^0\right) - C_{11} Z_{21} - C_{12} Z_{22} -
\gamma_0 C_{21},
\end{eqnarray}
with
\begin{eqnarray}
Z_{21}   &=& C\left(\frac{a^2}{\ve}\right) - \gamma_0 c_{10}\\
Z_{22}   &=& C\left(\frac{a^2}{\ve^{2}}\right)\\
C_{11}   &=& C\left(a \ve\right)   \\
C_{12}   &=& C\left(a \ve^2 \right)   \\
C_{21}   &=& C\left(a^2 \ve\right) - \gamma_0 C_{12}~.      
\end{eqnarray}
Furthermore the relations
\begin{eqnarray}
Z_{31} + Z_{11} c_{20} + Z_{22} C_{11} + Z_{21} c_{10} &=& 0 \\
Z_{32} + c_{10} Z_{22} &=& 0
\end{eqnarray}
hold. The above relations yield the anomalous dimensions and the Wilson coefficients
to $O(a^3)$.
\section{Results}

\vspace{1mm} \noindent
The 16th moment of the 3--loop non--singlet anomalous dimension 
$\gamma^{16,+}_{{\rm NS}}$, 
which describes the evolution of the combination
\begin{equation}
q^+_{\rm NS}(x,Q^2) = \left(q_j(x,Q^2) + \overline{q}_j(x,Q^2)\right) - \left(q_k(x,Q^2) + 
\overline{q}_k(x,Q^2)\right) 
\end{equation}
of quark densities, 
and the coefficient functions $C_{{2,L}, 
16}^{{\rm NS}}$ were calculated using the {\tt MINCER} algorithm 
\cite{MINCER}. 
The calculation was performed majorly using two dual--processor 32bit PC's 
(3 and 2.6 GHz). Several diagrams were run on an Opteron 64bit PC. Due 
to the large disk--space required by a series of diagrams a 4.2 Tbyte raid 
system was linked to the two 32bit PC's to store intermediary results. In 
the calculation the moment (\ref{eq19}) is determined in terms of an
$\varepsilon$--expansion for $a_s = a_0$. The anomalous dimension and 
moments of the coefficient functions are determined as described in the 
previous section. 

Up to 3--loop order 388 diagrams contribute effectively, if symmetry 
relations between diagrams are used. The calculation using the above system needed 
about 560 CPU days. The CPU time distribution over the individual 
diagrams is shown in Figure~1. Several diagrams required computation times 
of a month, in one case of $O(60)$ days. Despite of these long computation 
times the system ran completely stable. 
The use of a parallel facility running {\tt FORM}~\cite{PARFORM} would be 
highly desirable for computations of similar size in the future.

The 16th moments of the non--singlet anomalous dimension and coefficient 
functions for unpolarized nucleons are~:
\renewcommand{\arraystretch}{1.5}
\begin{eqnarray}
\gamma^{16,(0),+}_{\rm NS} &=& {\frac {64419601}{6126120}}\,{C_F} = 14.02075071 \\
& & \nonumber\\ 
\gamma^{16,(1),+}_{\rm NS} &=&
{\frac {
21546159166129889}{484994628518400}}\,{C_F}\,{C_A}
- {\frac {
3689024452928781382877}{459818557352009856000}}\,C_F^{2}\nonumber\\
\nonumber\\ & &
- \frac{1176525373840303}{112588038763200} C_F N_F
\nonumber\\ & & \nonumber\\ 
&=& 163.4395247 - 13.93310085~N_F
\\
& &\nonumber\\
\gamma^{16,(2),+}_{\rm NS} &=& - \left (
{\frac
{58552930270652300886778705063429867}{3451337970612452534317096673280000}}
- {\frac {59290512768143}{1563722760600}}\,{\zeta_3}
\right ) C_F^{3}
\nonumber\\ & &
\nonumber\\ & &
+\Biggl({\frac {1670423728083984207878825467}{
6488959481351563087872000}}+{\frac {59290512768143}{3127445521200}}\,{
\zeta_3}
\Biggr ){C_F}\,C_A^{2}
\nonumber\\ & &
\nonumber\\ & &
+\left (-{\frac {
1229794646000775781127856064477}{30335885575318557435801600000}}-{\frac {
59290512768143}{1042481840400}}\,{\zeta_3}\right ) C_F^{2}{ 
C_A}
\nonumber\\ & &
\nonumber\\ & &
+\left
(-{\frac {71543599677985155342551355451}{938967886855098206346240000}}+{\frac
{64419601}{765765}}\,{\rm \zeta_3}\right ) C_F^{2}{N_F} \nonumber\\  
& &+\Biggl (-{\frac {15018421824060388659436559}{
579371382263532418560000}} -
{\frac {64419601}{765765}}\,{\zeta_3}\Biggr ){
C_F
}\,C_A\,N_F 
\nonumber\\ 
& &
-{\frac {5559466349834573157251}{
2069183508084044352000}}\,{C_F}\, N_F^{2}
\nonumber\\ & &
\nonumber\\
&=&         2849.5632736921273714 
         - 463.86001156801831223~N_F
\nonumber\\ & &
         - 3.5823897546153993659~N_F^2~.
\end{eqnarray}
\begin{eqnarray}
& & \nonumber\\
C_2^{\rm NS,16}(a_s) &=& 1 + {\frac {4047739719}{190590400}}\,{ C_F}\,a_s 
\nonumber\\&+&
\Biggl[  \Biggl( {\frac {
44426674163044428879366970127}{321931846921747956461568000}}
+ {\frac {
24439538}{255255}}\,{ \zeta_3} \Biggr)  C_F^{2}
\nonumber\\ &+&
\Biggl( {\frac {
17918308408498294222783087}{59422705873182812160000}}-{\frac {
113298677}{1021020}}\,{  \zeta_3} \Biggr) {  C_F}\,{  C_A}
\nonumber\\ &-&
{\frac {
143568372761907472111177}{2758911344112059136000}}\,{  C_F}\,{  N_F}
 \Biggr] {a_s}^{2}
\nonumber\\ &+&
\Biggl[  \Biggl( 
{\frac {
3036813397599509725084677293842505976559161689}{
8034458016040775933421647863403347968000000}}
\nonumber\\ &+&
{\frac {
1494341926940450865387403}{595674040206012768000}}\,{  \zeta_3} 
+
{\frac {59290512768143}{3127445521200
}}\,{  \zeta_4}-{\frac {27643576}{21879}}\,{  \zeta_5}
\Biggr)  C_F^{3}
\nonumber\\ &+&
 \Biggl( 
{\frac {262865377883475726558800935515033190333}{
56646805852503848671021043712000000}}
-{\frac {15355050469171482313}{4991403051835200}}\,{  \zeta_3}
\nonumber\\ &+&
{\frac {59290512768143}{6254891042400}}\,{  \zeta_4}
+
{\frac {47187263}{51051}}\,{  
\zeta_5}
 \Biggr) {  C_F}\, C_A^{2}
\nonumber\\ &+&
\Biggl( {\frac {7750026627118768752845091760890051465242741}{
1652500620329242273431025887166464000000}}
\nonumber\\ &-&
{\frac {
2849482004138921491531}{6741167121672984000}}\,{  \zeta_3}
- {\frac {59290512768143}{2084963680800}}\,{  \zeta_4}
+{\frac {983963
}{21879}}\,{  \zeta_5}
 \Biggr)   C_F^{2}{C_A}
\nonumber\\ &+&
\Biggl( 
{-\frac {
4073207241348493196152222079933557529}{
3529777469944553728278848870400000}}
-{\frac {552298563960959}{
4021001384400}}\,{  \zeta_3}
\nonumber\\ &+&
{\frac {64419601}{1531530}}\,{
  \zeta_4} \Biggr)   C_F^{2}{  N_F}
\nonumber\\ &+&
\Biggl(
-{
\frac {582811634921542995647179358698536547}{
404620041803598919078721740800000}} 
+
 {\frac {598788865585667
}{1850495446800}}\,{  \zeta_3}
\nonumber\\ &-&
{\frac {64419601}{1531530}}\,{  \zeta_4}
\Biggr) {  C_F}\,{  C_A}\,{  
N_F} 
\nonumber\\ &+&
 \Biggl( {\frac {
7227384935999670312318789884999}{76056398835262954714045440000}}+{
\frac {64419601}{20675655}}\,{  \zeta_3} \Biggr) {  C_F}\,  N_F^{2}
\nonumber\\ &+& \langle e \rangle
\left( 
{\frac {705894258514655486993}{3248429831350704000}} 
+{\frac {38404365803}{1533061530}}\,{\zeta_3}
-{\frac {14560}{51}}\,{\zeta_5}
\right) \frac{d_{abc}^2}{N_c} N_F
\Biggr] {a_s}^{3}
\nonumber\\ & &
\nonumber\\ & &
\nonumber\\  &=& 1 + 28.31719904~a_s + (1122.549565 - 69.38406971~N_F)~a_s^2
\nonumber\\  & &
     + (50309.36422 - 6651.875513~N_F + 131.6959033~N_F^2 
\nonumber\\ & & -216.0757466~\langle e \rangle N_F 
 )~a_s^3
\end{eqnarray}

\begin{eqnarray}
& &\nonumber\\
C_L^{\rm NS,16}(a_s)
&=&
{\frac {4}{17}}\,{C_F}\,a_s
\nonumber\\ &+&
\Biggl[ \left( -\frac {29393927457809}{
44659922042736} + \frac{96}{17} \zeta_3 \right)
\,C_F^{2} 
+ \left(
{\frac {
55969347000169}{8209544493150}}  - {\frac {48}{17}}\,{\zeta_3}\right)\,{C_F}\,{C_A}
\nonumber\\  &-&
{\frac {39366889}{39054015}}\,{ C_F
}\,{N_F}
\Biggr] {a_s}^{2}
\nonumber\\ &+&
\Biggl[  \Biggl(
-\frac{7508281821276771498126447290110919}{
13647898235438852429242598400000} 
-{\frac {196256899828170631}{133698296031300}}\,{\zeta_3}
\nonumber\\ &+&
{\frac {39360}{17}}\,{\zeta_5}
\Biggr) C_F^{3}
\nonumber\\ &+&
 \Biggl( {
\frac {296045501010133565322039207159677}{
936620467137960460830374400000}}
+ {\frac
{2253147763389895}{1188429298056}}\,{\zeta_3} 
\nonumber\\ &-&
{\frac {40160}{17}}\,{\zeta_5} 
\Biggr) {C_A}\,C_F^{2}
\nonumber\\ &+&
\Biggl( 
{\frac {1460792499427100139493280371}{
8256042197255336964480000}}
- {\frac {1634895686765221}{
2673965920626}}\,{\zeta_3}
\nonumber\\ &+&
{\frac {10240}{17}}\,{\zeta_5} \Biggr)
C_A^{2}{C_F} 
\nonumber\\ &+&
\Biggl( {\frac {3529137346321170453160463}{
136796020812222932160000}} - {\frac {44651224}{765765}}\,{\zeta_3}
 \Biggr) C_F^{2}\, N_F
\nonumber\\ &+&
\Biggl( - {\frac {4495805144658565385501573689}{
57792295380787358751360000}}
+
{\frac {43594330672}{1249937325}}\,{\zeta_3} 
\Biggr) {C_A}\,{C_F}\,N_F \nonumber\\ &+& 
{\frac {895967716232}{209134250325}}\,{C_F}\,N_F^{2}
\nonumber\\ &+&  \langle e \rangle
\left(-{\frac {1798450729620489619601}{18272417801347710000}}
-{\frac {28854977192}{547521975}}\,{\zeta_3}+{\frac {2560}{17}}
\,{\zeta_5}
 \right) \frac{d_{abc}^2}{N_c} N_F
\Biggr] {a_s}^{3}
\nonumber\\ 
&=& 0.3137254902~a_s + (24.59183098 - 1.344015086~N_F)~a_s^2
\nonumber\\ & &     + (1779.007454 - 222.2146165~N_F + 5.712233265~N_F^2 
\nonumber\\ & &
- 24.99942737~\langle e \rangle N_F)~a_s^3~,  
\end{eqnarray}

\noindent
with 
\begin{eqnarray}
\langle e \rangle = \frac{3}{N_F} \sum_{k=1}^{N_F} e_f
\end{eqnarray}
and $d_{abc}^2/N_c = 40/9$ for $SU(3)_c$.
These results agree with the complete 3--loop results for the anomalous 
dimension in~\cite{3lns}. The moments for the $N_F^2$--terms \cite{GRAC} 
and $N_F$--terms \cite{NFNS} were known before. For the coefficient 
functions the moments agree with the very recent complete results 
\cite{3lFL} and an upcoming paper \cite{3lcoef}.   

\section{Conclusions} 

\vspace{1mm}
\noindent
We calculated the hitherto unknown 16th moments for the 3--loop 
non--singlet anomalous 
dimension $\gamma_{\rm NS}^{16,+}$ and the non--singlet coefficient functions $C_{2,L}(x,Q^2)$ for 
pure photon exchange. The computation was performed using the {\tt MINCER} 
algorithm, which is different from the algorithms used in the 
recent complete calculations. In view of the rather long CPU time of 
about 560 days spent 
for the calculation the reliability of the formula manipulation program 
{\tt FORM} has been tested intensely as a by--product. The results agree with recent and upcoming
complete results.
The computation of the 16th moment provides a  non--trivial test of 
these computations.

\vspace{2mm}\noindent
{\bf Acknowledgment.} 
We would like to thank S. Moch for useful discussions, U. Gensch for 
support of the project and S. Wiesand, P. Wegner and C. Spiering for 
their technical support. This paper was supported in part by DFG 
Sonderforschungsbereich Transregio~9, Computergest\"utzte Theoretische
Physik.


\vspace*{-1cm}
\begin{center}

\mbox{\epsfig{file=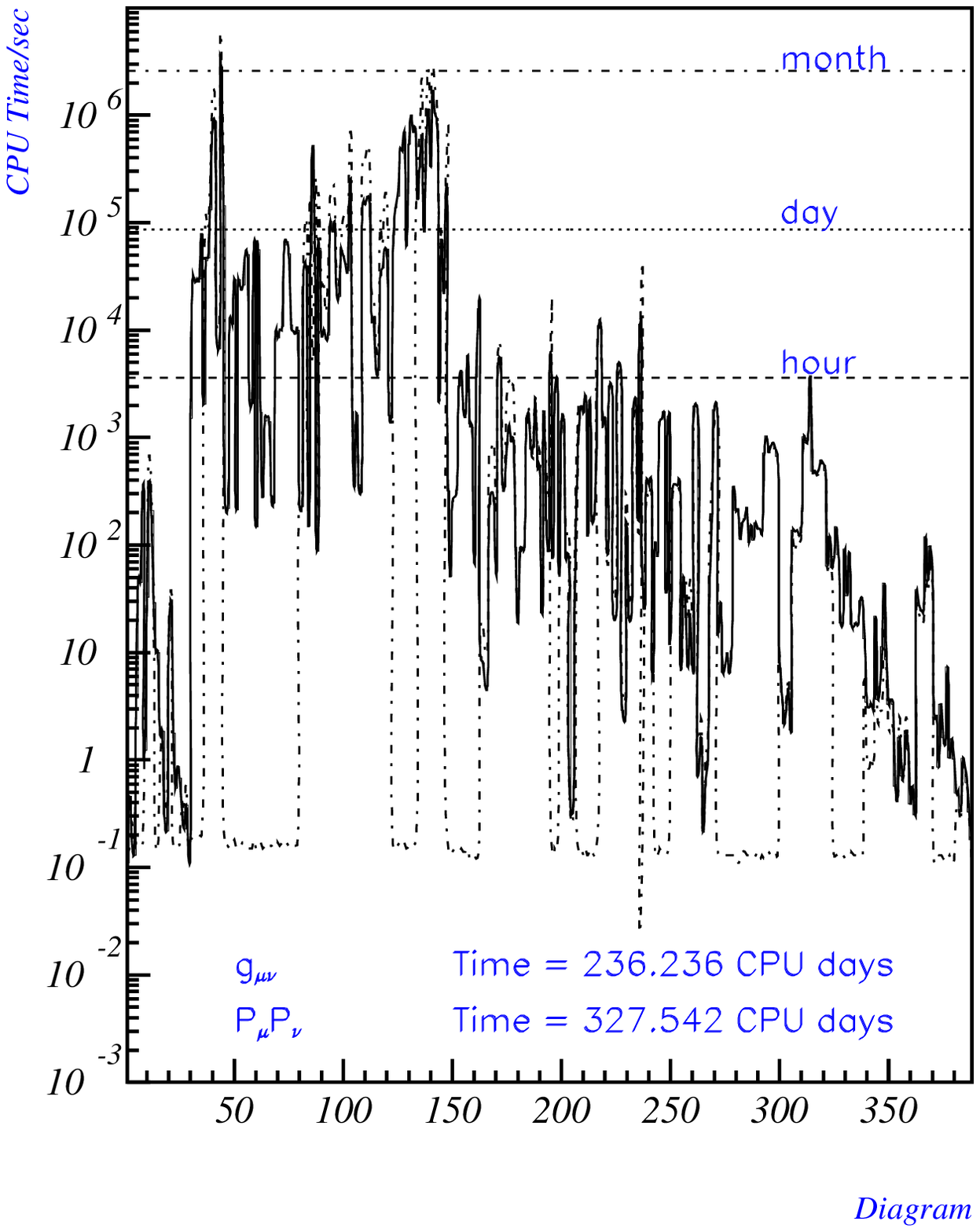,height=16cm,width=16cm}}

\vspace{2mm}
\noindent
\small
\end{center}
{\sf
Figure~1:~Execution time profile for the set of diagrams. Full line:
$g_{\mu\nu}$ projection; dash-dotted line: $P_{\mu} P_{\nu}$ projection.
\normalsize


\newpage
\begin{thebibliography}{10}
%
\bibitem{H1Z}
M. Botje, M. Klein, and C. Pascaud, in~:
Workshop on Future Physics at HERA Hamburg, Germany, 30-31 May 1996, pp.~33, eds. G. 
Ingelman, A De Roeck, and R. Klanner,
{\tt hep-ph/9609489}.
%
\bibitem{NLOPR}
J. Bl\"umlein, S. Riemersma, W.L. van Neerven, and A. Vogt, Nucl. Phys. {\bf B} Proc. 
Suppl. {\bf 51C} (1996) 97, {\tt hep-ph/9609217}.
%
\bibitem{BBG}
J. Bl\"umlein, H. B\"ottcher, and A. Guffanti, {\tt hep-ph/0407089}.
%
\bibitem{BETH}
S. Bethke, {\tt hep-ex/0407021}.
%
\bibitem{gLO}
D.J. Gross and F. Wilczek,
Phys. Rev. {\bf D8} (1973) 3633; {\bf D9} (1974) 980;\\
H. Georgi and H.D. Politzer, Phys. Rev. {\bf D9} (1974) 416.
%
\bibitem{FP}
W. Furmanski and  R. Petronzio, Z. Phys. {\bf C11} (1982) 293 and references therein.
%
\bibitem{gNLO}
E.G. Floratos, D.A. Ross, and C.T. Sachrajda, Nucl. Phys. {\bf B129}  (1977) 66;
{\bf B139} (1978) 545; {\bf B152} (1979) 493;\\
A.~Gonzalez-Arroyo, C.~Lopez, and F.J. Yndurain, Nucl. Phys. {\bf B153}(1979) 
161;\\
A.~Gonzalez-Arroyo and C.~Lopez, Nucl. Phys. {\bf B166} (1980) 429;\\
C.~Lopez and F.J. Yndurain, Nucl. Phys. {\bf B183} (1981) 157;\\
W.~Furmanski and R.~Petronzio, Phys. Lett. {\bf B97} (1980) 437;\\
G.~Curci, W.~Furmanski, and R.~Petronzio, Nucl. Phys. {\bf B175} (1980) 27;\\
R.~Hamberg and W.~L. van Neerven, Nucl. Phys. {\bf B379} (1992) 143;\\
R. Mertig and W.L. van Neerven, Z. Phys. {\bf C70} (1996) 637;\\
W. Vogelsang, Phys. Rev. {\bf D54} (1996) 2023.
%
\bibitem{cNLO}
D.W. Duke, J.D. Kimel, and G.A. Sowell, Phys. Rev. {\bf D25} (1982) 71;\\
A.~Devoto, D.W. Duke, J.D. Kimel, and G.A. Sowell, Phys. Rev. {\bf D30} (1984) 541;\\
D.I. Kazakov and A.V. Kotikov, Nucl. Phys. {\bf B307} (1988) 721; Phys. Lett. {\bf 
B291}  (1992) 171;\\
D.I. Kazakov, A.V. Kotikov, G.~Parente, O.A.~Sampayo, and J.~Sanchez 
Guillen Phys. Rev. Lett. {\bf 65} (1990) 1535;\\
J.~Sanchez Guillen, J.~Miramontes, M.~Miramontes, G.~Parente, and 
O.A. Sampayo, Nucl. Phys. {\bf B353} (1991) 337;\\  
W.L. van Neerven and E.B. Zijlstra, Phys. Lett. {\bf B272} (1991) 127;\\
E.B. Zijlstra and W.L. van Neerven, Phys. Lett. {\bf B273} (1991) 476; Nucl. Phys. 
{\bf B383} (1992) 525;\\
S.A. Larin, J.A.M. Vermaseren, Z. Phys. {\bf C57} (1993) 93;\\
S. Moch and J.A.M.~Vermaseren, Nucl. Phys. {\bf B573} (2000) 853;\\
W.L. van Neerven and E.B. Zijlstra,
Nucl.Phys. {\bf B417} (1994) 61, E: {\bf B426} (1994) 245.
%
\bibitem{NNLOmom}
S.A. Larin, F.V. Tkachov, J.A.M. Vermaseren, Phys. Rev. Lett. {\bf 66} 
(1991) 862;
Phys. Lett. {\bf B272} (1991) 121;\\
S.A. Larin and J.A.M. Vermaseren, Phys. Lett. {\bf B259}  (1991) 345;\\
S.A. Larin, T.~van Ritbergen, and J.A.M. Vermaseren,
Nucl. Phys. {\bf B427} (1994) 41;\\
S.A. Larin, P.~Nogueira, T.~van Ritbergen, and J.A.M. Vermaseren,
Nucl. Phys. {\bf B492} (1997) 338.
%
\bibitem{FORM}
J.A.M. Vermaseren, 
{\em Symbolic Manipulation with {\sc Form} version 2, Tutorial and
  Reference Manual} (Computer Algebra Nederland, Amsterdam, 1991); 
{\em New Features of FORM} {\tt  math-ph/0010025}.
%
\bibitem{MOM14}
A. Retey and J.A.M. Vermaseren, Nucl. Phys. {\bf B604} (2001) 281.
%
\bibitem{3lns}
S. Moch,  J.A.M. Vermaseren, and A. Vogt, Nucl. Phys. {\bf B688} (2004) 
101.
%
\bibitem{3lsin}
A. Vogt, S. Moch,  and J.A.M. Vermaseren, Nucl. Phys. {\bf B691} (2004) 
129. 
%
\bibitem{MINCER}
S.A. Larin, F.V. Tkachev, and J.A.M. Vermaseren,~NIKHEF-H-91-18.
%
\bibitem{3lFL}
S. Moch,  J.A.M. Vermaseren, and A. Vogt, {\tt hep-ph/0411112}.
%
\bibitem{3lcoef}
S. Moch,  J.A.M. Vermaseren, and A. Vogt, in preparation. 
%
\bibitem{LCE}
K.G.~Wilson, Phys. Rev. {\bf 179} (1969) 1699;\\
R.A.~Brandt and G.~Preparata, Fortschr. Phys. {\bf 18} (1970)
249;\\
W.~Zimmermann, {\sf Lect. on Elementary Particle Physics and
Quantum
Field Theory}, Brandeis Summer Inst., Vol.~1,
(MIT Press, Cambridge, 1970),~p. 395;\\
Y.~Frishman, Ann. Phys. {\bf 66} (1971) 373.
%
\bibitem{BET}
D.J. Gross and F. Wilczek, Phys. Rev. Lett. {\bf 30} (1973) 1343;\\
H.D. Politzer, Phys. Rev. Lett. {\bf 30} (1973) 1346;\\
D.R.T. Jones, Nucl. Phys. {\bf B75} (1974) 531;\\
W.E. Caswell, Phys. Rev. Lett. {\bf 33} (1974) 244;\\
O.V. Tarasov, A.A. Vladimirov, and A.Yu. Zharkov, Phys. Lett. {\bf 93B} (1980) 429;\\
S. Larin and J.A.M. Vermaseren, Phys. Lett. {\bf B303} (1993) 334.
%
\bibitem{PARFORM}
D.~Fliegner, A.~Retey, and J.A.M. Vermaseren,
{\tt hep-ph/9906426}.
%
\bibitem{GRAC}
J.~A. Gracey, Phys. Lett. {\bf B322}   (1994) 194.
%
\bibitem{NFNS}
S. Moch, J. Vermaseren, and A. Vogt, Nucl. Phys. {\bf B646} (2002) 181. 
\end{thebibliography}
\end{document}